\documentclass{osa-article}
\journal{oe}
\articletype{Research Article}
\usepackage{multirow,tabularx}
\newcommand{\doi}[2]{\href{https://doi.org/#1}{#2}}

\begin{document}

\title{Optical characterisation of micro-fabricated Fresnel zone plates for atomic waveguides}

\author{V.~A.\ Henderson,\authormark{1,2,*} M.~Y.~H.\ Johnson,\authormark{1} 
Y.~B.\ Kale,\authormark{1,3}  P.~F.~Griffin,\authormark{1} E.\ Riis,\authormark{1} and A.~S.\ Arnold\authormark{1,$\dagger$}}

\address{\authormark{1}Dept.\ of Physics, SUPA, University of Strathclyde, Glasgow, G4 0NG, UK\\
\authormark{2}Now at Dept.\ of Physics, Humboldt-Universit\"at zu Berlin, 12489 Berlin, Germany\\
\authormark{3}Now at Dept.\ of Physics, University of Birmingham, Birmingham, B15 2TT, UK\\
\email{\authormark{*}henderson@physik.hu-berlin.de} 
\email{\authormark{$\dagger$}aidan.arnold@strath.ac.uk}}

\begin{abstract}
	We optically assess Fresnel zone plates (FZPs) that are designed to guide cold atoms. Imaging of various ring patterns produced by the FZPs gives an average RMS error in the brightest part of the ring of 3\% with respect to trap depth. This residue will be due to the imaging system, incident beam shape and FZP manufacturing tolerances. Axial propagation of the potentials is presented experimentally and through numerical simulations, illustrating prospects for atom guiding without requiring light sheets.
\end{abstract}

\section{Introduction}

Inertial and rotation sensing are important both for everyday navigation and the exploration of fundamental Physics. Highly precise, accurate and stable devices are of particular interest for inertial navigation, geodesy, geophysics, and tests of general relativity~\cite{Cronin2009,Schreiber2013,Barrett2014}. For gyros the most sensitive results have been seen with atomic beams \cite{Gustavson2000,Durfee2006} and $16\,$m$^2$ area ring lasers \cite{Schreiber2011}.   
The use of cold atoms rather than light has the potential to create higher sensitivity -- and crucially, lower drift -- devices which could operate either alone or in a hybrid classical-quantum system~\cite{Cronin2009,Barrett2014}. Despite this potential, the practicality of quantum technologies, such as rotation sensing, are contingent on reducing the footprint of existing ultracold atom technologies -- which may be aided by new trapping and cooling techniques \cite{McGilligan2015, McGilligan2017,Barker2019,Kang2019}. 

The sensitivity of standard atom interferometers depends on the square of the interferometer time or the area enclosed (depending on the configuration) and these are therefore key parameters to maximise. For both light and atoms the sensitivity is also dependent on the squareroot of particle flux, and the measurement rate determines the dynamic range. Existing high-end atom interferometric sensors use a variety of techniques to work towards these enhancements, including fountains~\cite{Dickerson2013, Dutta2016}, `juggled' or exchanged atom clouds~\cite{Rakholia2014, Dutta2016}, micro-gravity~\cite{Muntinga2013, Liu2018, Becker2018, vicki2019},  and waveguiding~\cite{Burke2009, Marti2015}. 

Atom guiding is of particular interest to us as it has the potential to reduce the size of experiments from tens of meters, to something even smaller than the portable `fountain' systems already demonstrated \cite{Leykauf,landragin2018}. While ring-guided ultracold atom interferometers have long been a goal \cite{Arnold2006,Marti2015}, actual rotation metrology has only been achieved very recently \cite{sackett2019,boshier2019}. Atom waveguides, in particular rings, are also of great interest outside inertial sensing for a variety of fundamental physics experiments including dimensionality~\cite{Ville2017, Aidelsburger2017}, cosmological phenomena~\cite{Eckel2018a}, and superfluidity~\cite{Eckel2014}.

In response to this, we explore the suitability of Fresnel zone plates (FZPs) for the production of optical waveguides for cold atoms, with a view toward compact large-enclosed-area interferometric devices. Utilising high precision microfabrication, FZPs are exciting candidates for the production of static trapping potentials useful to atomtronics \cite{atomtronicanderson,Eckel2014}, interferometry, and fundamental physics. They are particularly suitable for compact quantum technologies due to their simplicity, planar form factor, and the potential for low cost mass-production.  In this paper, we present the optical testing of a variety of Fresnel zone plate geometries -- yielding high-intensity rings, guides and beamsplitters -- that were manufactured following our theoretical proposal \cite{Henderson2016}.

We image each of the potentials produced by the zone plates, with methods and preliminary results detailed in Sec.~\ref{sec:results}.  We then analyse the ring potentials in detail in Sec.~\ref{sec:analysis}: from basic characterisation of the ring geometry and efficiency in Sec.~\ref{sec:ringchar} to ring roughness analysis in Sec.~\ref{sec:smoothness}. An RMS error approach is used initially, followed by azimuthal Fourier analysis of the potentials to determine the dominant length scales involved. 

The kinoforms characterised in this paper were designed according to the methods outlined in Refs.\ \cite{Henderson2016, McDonald2015,Zhai2017}. Briefly, we utilise a Fourier-optics method of modelling the propagation of an electric field \cite{ifancharles}, which allows us to analytically calculate the phase map of the kinoform from a target intensity pattern in the focal plane. This procedure allows us to create kinoforms of any phase depth, and with spatial resolution of around a wavelength, without using any optimisation algorithms commonly used for kinoform generation.  A selection of 24 patterns, with $2\,$mm separation and $2\,$mm diameter, were etched with regularly spaced $1\times 1\,\upmu$m$^2$ `pixels' onto a $18\times 26 \times 3\,$mm$^3$ fused silica substrate by Holo/Or \cite{Kedmi1990}. The binary patterns have two distinct heights resulting in a modulation of the propagated light phase by $\pi$ between zones of the FZP \cite{Henderson2016}. For shorter optical wavelengths, and $10\,$nm  resolution over cm$^2$ areas, electron-beam lithography could be used for FZP fabrication \cite{nshii2013}.

\begin{figure}[!b]
	\centering
	\includegraphics{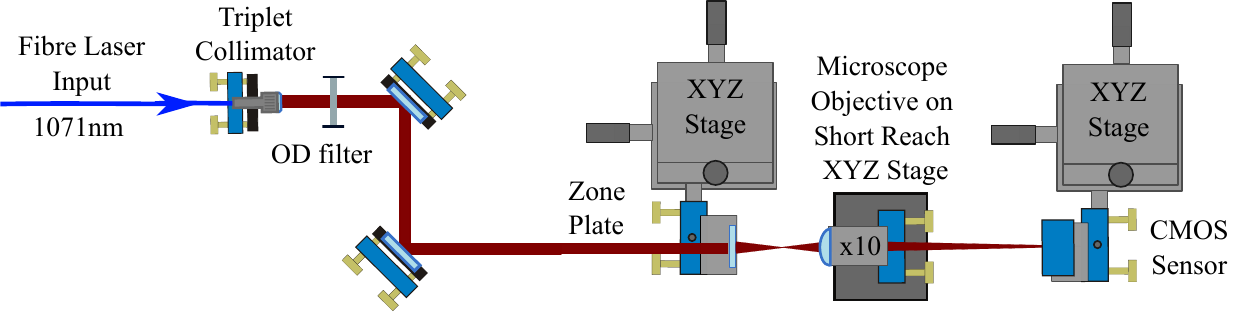} 
	\caption[Manufactured zone plates]{\label{fig:Schematic} The optical layout used to image the focal plane of the FZPs.}
\end{figure}

The patterns were designed for the wavelength of our dipole laser, $\lambda_\textrm{D}=1071\,$nm, which was chosen to be red-detuned from the nearest (D$_1$ and D$_2$) rubidium lines at $795\,$nm and $780\,$nm. Cold atoms released into the light would therefore be trapped in the brightest parts of the pattern as the optical dipole potential is proportional to the intensity and the reciprocal of detuning. Blue-detuned potentials, where atoms are attracted to regions of darkness, are interesting as heating due to intensity-dependent light scattering is dramatically reduced \cite{arnold2012,birkl}. We consider the extension of FZPs to this realm in future work \cite{fzp3}.

\section{Experimental setup and FZP pattern overview} \label{sec:results}

Optical testing was carried out using the layout shown in Fig.~\ref{fig:Schematic}, which consisted of three main sections: incident beam preparation, zone plate mounting, and imaging. 
In the preparation section, a collimated Gaussian beam with a $w_\textrm{D}=1\,$mm $1/e^2$ radius and $M^2$ close to one was created.
As our maximum FZP numerical aperture (NA) was only 0.05-0.14, linearly polarised light was used, however for higher NA FZPs the transverse spatial polarisation will need to be cylindrically symmetric to avoid an asymmetric focus \cite{tightfocus}.

In our optical tests we are using an unamplified kHz-linewidth seed laser. For atom trapping, several Watts of optical power will be required. Lasers with these kind of powers can suffer from pointing stability, as well as relative intensity noise (RIN) at the percent level. Pointing stability issues are largely removed with appropriate (high $M^2$) fiber amplifiers, and an acousto- or electro-optic based noise-eater can also be used to dramatically reduce RIN over a wide frequency range. 

The imaging section of the system consists of a $\times10$ microscope objective (with a numerical aperture of $0.25$)  and a CMOS-sensor camera (Cinogy CMOS1201). Both of these components are mounted on translation stages and kinematic mounts to allow the precise alignment required to minimise system aberrations.

The zone plate is mounted in a standard prism mount and attached to a three-dimensional translation stage, giving full control of the zone plate's position and orientation relative to the incident beam.  The horizontal and vertical ($XY$) translation directions are used to select the zone plate pattern, and the direction parallel to the incident beam ($Z$) is used to scan the focus of the zone plate through the imaging plane of the objective lens. This method of longitudinal scanning ensures that the effective focal length and magnification of the imaging system stay constant for all zone plates.

\begin{figure}[!t]
	\centering
    \includegraphics[width=\textwidth]{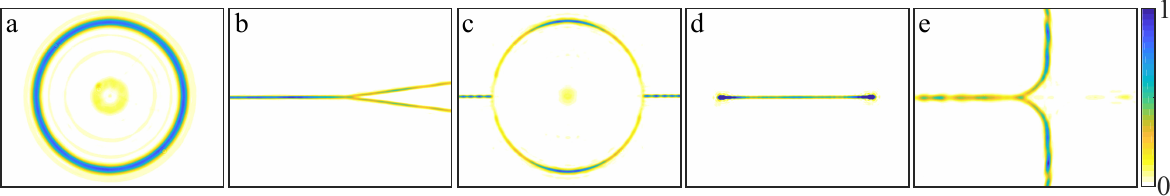}
	\caption[Experimental images of zone plate patterns]{\label{fig:rawdata} Experimental images of five example FZPs averaged over 3-7 individual shots. Acquisition conditions are discussed in the text. Each plot has area $590\times480\,\upmu\textrm{m}^2$, and the colour bar indicates relative light intensity.}
\end{figure}

The imaging system magnification is calibrated via comparison with ring images obtained directly on the sensor without a magnifying lens, which confirmed the ring radii exactly matched the design values. This calibration method gives a mean magnified resolution of $0.464(7)\,\upmu$m per pixel.
Magnified experimental images of five example patterns are shown in Fig.~\ref{fig:rawdata} (the 24 etched patterns are variations on these five patterns). The images are captured at the FZP focus and with the maximum camera exposure possible without over-exposure.  An exception to this is the line pattern (Fig.~\ref{fig:rawdata} d), which is over-exposed so that detail in the middle of the line can be seen, as discontinuities at the ends of the line create very bright spots which overwhelm the rest of the pattern.  Each of the plots show an area of $1280\times1024$ pixels, taken after the objective lens. 

The patterns shown in Fig.~\ref{fig:rawdata} show good reproduction of general shapes, even for the complex case of the `ring-track' (Fig.~\ref{fig:rawdata} c), however it is obvious that many of the potentials do not have the smooth minima that would be required to operate successfully as a waveguide for ultracold atoms.  Due to their circular symmetry, the rings (Fig.~\ref{fig:rawdata} a) look very smooth, but all the other, less symmetric patterns have significant intensity variations in the guiding direction, where atoms will flow. For example, the addition of input/output couplers to a ring, making a ring track, creates a discontinuity at the junctions and removes the circular symmetry present in one of the isolated ring traps. Note also that results are for Gaussian beam illumination, rather than the ideal spatial intensity distribution for each FZP. Ideal  illumination, using spatially shaped intensity and phase, can be accurately achieved using spatial light modulator techniques \cite{Clark2016,Offer2018}, and will be considered in more detail in Ref.~\cite{fzp3}.

To establish the suitability of our zone plates for atom trapping and guiding, we characterise the ring potentials in terms of geometry (Sec.~\ref{sec:ringchar}) and smoothness (Sec.~\ref{sec:smoothness}). 

\section{Analysis and discussion} \label{sec:analysis}

\subsection{Ring parameters and scaling laws}


We characterise rings by taking images and  `unfolding' them, i.e.\ going from Cartesian to polar co-ordinates using angular resolution corresponding to one camera pixel at the ring radius. We can then individually fit these radial slices to a Gaussian with no DC offset, as the images are background subtracted. We only fit to the portion of the intensity which is well approximated by a Gaussian (the top 70\%) so as to exclude variation in the low intensity regions which are irrelevant for sufficiently cold guided atoms. Full 2-dimensional ring fits are possible if required. The target intensity profile at the focus of the FZP has the cylindrically symmetric form:
\begin{equation}
    I=I_0 \exp\left( \frac{-2 (r-r_0)^2}{w^2}\right),\;\;\;\textrm{with}\;\;\; I_0=\frac{P_\eta }{2^{1/2} \pi^{3/2} r_0 w}\approx 0.127\frac{P_\eta}{r_0 w},\label{eq:ringeq}
\end{equation}
where $I_0$ is the ring peak intensity at the ring radius $r_0$,  and $w$ is the waist or $e^{-2}$ intensity half-width. A beam with power $P$ incident on the FZP is focused with efficiency $\eta$ into the ring -- i.e.\ the effective power is $P_\eta=\eta\, P$.

From the ring spatial intensity a variety of other experimentally relevant parameters and their scaling laws can be determined. The detailed comparison specifically for Rb and at our dipole wavelength is provided in the Appendix, however, here we present a summary parameter dependence on $P_\eta,$ $r_0$ and $w$:
\begin{equation}
T\propto R \propto \frac{P_\eta}{r_0\, w}, \;\;\;
\nu_r \propto  \frac{{P_\eta}^{1/2}}{{r_0}^{1/2}\, w^{3/2}}, \;\;\;
\nu_z \propto  \frac{\lambda_\textrm{D}{P_\eta}^{1/2}}{{r_0}^{1/2}\, w^{5/2}}, \;\;\;
a_\textrm{max} \propto  \frac{P_\eta}{r_0\, w^2}, 
\label{scaling}
\end{equation} 
where $T,$ $R$, $\nu_r,$ $\nu_z,$ and $a_\textrm{max}$ are the potential depth (in temperature), the photon scattering rate, the radial and axial harmonic trap frequencies, and the maximum radial acceleration provided by the ring, respectively. 

\subsection{Geometry characterisation} \label{sec:ringchar}

\begin{figure}[!b]
	\centering	
    \includegraphics{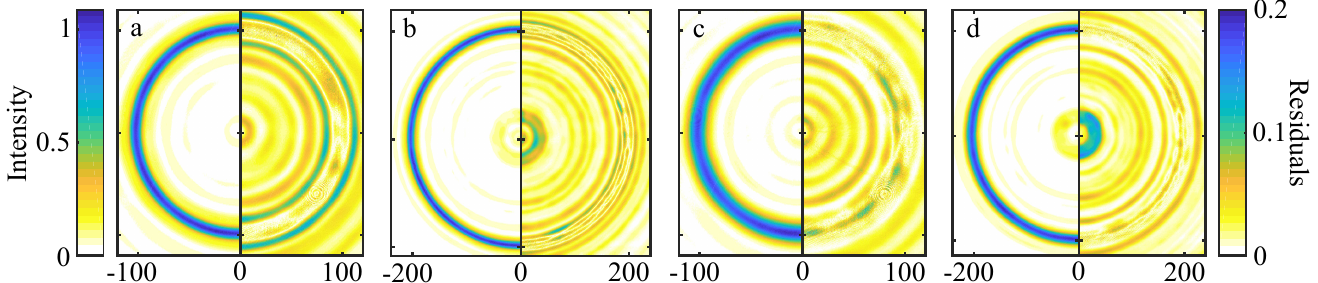} 
	\caption[Magnified rings with fitting]{\label{fig:rings} Magnified experimental images of the first 4 rings detailed in Table~\ref{tab:ringparams}. For each ring, the left side shows intensity (normalised by mean peak intensity), and residuals compared to the model in Eq.~\ref{eq:ringeq} are shown on the right. Horizontal distance is in microns.}
\end{figure}

Well-reproduced experimental ring intensities were observed with no perceivable ellipticity and radii matching their design parameters. Images of example rings and their respective fitted residuals are shown in Fig.~\ref{fig:rings} (see the dataset \cite{dataset} for data from all rings), with trap parameters obtained for all magnified ring potentials shown in Table~\ref{tab:ringparams}.  The displayed trap parameters are calculated from averaged radial slice fits rather than global fits. 

 \begin{table}[!t]
	\begin{center}
			\small
		\begin{tabular}{|c|c|c|c|c|c|c|c|c|c|c|c|}
			\cline{1-12}
			\multicolumn{3}{ |c| }{\textbf{Design}} & \multicolumn{4}{ c| }{\textbf{Observed}} & \multicolumn{5}{ c| }{\textbf{Inferred}}\\ \cline{1-12}
			$f$ & $r_0$ & $w$ & $r_0$ & $w$ & $\eta\,(\%)$ & $E\,(\%)$ & $T$ & $E$ & $\nu_r$ & $R$ & $a$\\ \cline{1-12}
			\multicolumn{1}{ |c  }{\multirow{5}{*}{18.2} } &
			\multicolumn{1}{ |c| }{100} & 5 & 100(1) & 9.1(1) & 22(4) \textit{22(1)} & 3.01 & 63 & 1.8 & 2.7 & 0.4 & 82 \\ \cline{2-12}
			\multicolumn{1}{ |c  }{}                        &
			\multicolumn{1}{ |c| }{200} & 5 & 199(3) & 11.9(2) & 26(2) \textit{29(1)} & 3.77 & 24 & 0.9 & 1.3 & 0.2 & 24   \\ \cline{2-12}
			\multicolumn{1}{ |c  }{}                        &
			\multicolumn{1}{ |c| }{100} & 10 & 100(1) & 14.1(2) & 25(2) \textit{27(1)} & 4.17 & 41 & 1.7 & 1.4 & 0.3 & 34   \\ \cline{2-12}
			\multicolumn{1}{ |c  }{}                        &
			\multicolumn{1}{ |c| }{200} & 10 & 203(3) & 14.8(2) & 27(2) \textit{31(1)} & 3.12 & 19 & 0.6 & 0.9 & 0.1 & 15  \\ \cline{2-12}
			\multicolumn{1}{ |c  }{}                        &
			\multicolumn{1}{ |c| }{200} & 15 & 205(3) & 14.9(2) & 26(2) \textit{28(1)} & 3.87 & 19 & 0.7 & 0.9 & 0.1 & 15   \\ \cline{1-12}
			\multicolumn{1}{ |c  }{\multirow{3}{*}{10} } &
			\multicolumn{1}{ |c| }{100} & 5 & 100(1) & 7.0(1) & 25(3) & 3.26 & 82 & 2.7 & 4.0 & 0.5 & 138  \\ \cline{2-12}
			\multicolumn{1}{ |c  }{}                        &
			\multicolumn{1}{ |c| }{200} & 5 & 199(3) & 8.8(2) & 25(2) & 3.51 & 33 & 1.2 & 2.0 & 0.2 & 44 \\ \cline{2-12}
			\multicolumn{1}{ |c  }{}                        &
			\multicolumn{1}{ |c| }{200} & 10 &  195(3) & 9.0(2)  & 24(2) & 4.61 & 33 & 1.5 & 2.0 & 0.2 & 43 \\ \cline{1-12}
			\multicolumn{1}{ |c  }{\multirow{2}{*}{7} } &
			\multicolumn{1}{ |c| }{100} & 5 &  102(1) & 5.8(1)  & 23(3) & 3.66 & 97 & 3.6 & 5.3 & 0.6 & 197 \\ \cline{2-12}
			\multicolumn{1}{ |c  }{}                        &
			\multicolumn{1}{ |c| }{200} & 5 &  197(3) & 6.2(1) & 22(2) & 3.56 & 47 & 1.7 & 3.4 & 0.3 & 89 \\ \cline{1-12}
		\end{tabular}
		\caption{\label{tab:ringparams} Design and observed ring parameters: focal length $f$ in mm, ring radius $r_0$ and waist $w$ both in microns, efficiency $\eta$, RMS error $E$. The errors quoted correspond to the standard deviation of the trap parameters measured from slices around the ring. The efficiency shown is the percentage of power transferred from the incident beam into an area of interest determined by the ring parameters, with unmagnified ring values shown in italics. The RMS error is calculated between measurements and the top 10\% of an ideal ring of the same width, radius and average depth. Subsequent columns show the inferred resultant trap depth $T$ and RMS error $E$ converted to $\upmu$K units, radial trap frequency $\nu_r$ in kHz, scattering rate $R$ in Hz and maximum radial acceleration $a$ in units of $g=9.81\,\textrm{m}\,\textrm{s}^{-2}$ for all manufactured rings (using Eqs.~\ref{temp}, \ref{nur}, \ref{rate} and \ref{amax}, respectively).  The trap frequency/depth is calculated using the observed trap waist, assuming a $10\,$W incident beam and 30\% efficiency. The RMS error $E$ in $\upmu$K gives an indication on temperature bounds for propagation -- unless atoms are released at an azimuthal potential maximum.}
	\end{center}
\end{table}


When comparing the designed and measured trap width, it is clear that the focal length $f$ of the kinoform (and so the numerical aperture) greatly influences the trap width. The approximate paraxial diffraction limit for a beam with $1/e^2$ radius $w_\textrm{D}$, i.e.\ $w_\textrm{min} = \lambda f / (\pi w_\textrm{D})$ yields minimum ring widths $w_\textrm{min}=\{2.4,3.4,6.1\}\,\upmu$m for our wavelength, beam size, and the focal lengths we used, i.e. $f= \{7, 10, 18.2\}\,$mm, respectively. Empirically adding a multiple ($1.25\times$) of this diffraction `resolution' $w_\textrm{min}$ in quadrature with the $w=5\,\upmu$m design widths gives very similar total minimum widths, $w=\{5.9,6.6,9.2\}\,\upmu$m, to those observed in Table~\ref{tab:ringparams}, suggesting that the potentials are nearly diffraction limited, with a more detailed examination in Ref.~\cite{Henderson2018}. The theoretical design includes diffraction, however the designed focal resolution is contingent on the ideal spatial intensity at the FZP, rather than the Gaussian illumination used here.    

We define the efficiency $\eta$ of the FZPs as the percentage of incident power which is directed into an area of interest relevant to the atoms, i.e.\  within $3w$ of the ring radius $r_0$ using the fitted Eq.~\ref{eq:ringeq}. The results are shown in Table~\ref{tab:ringparams}, with the italic values indicating unmagnified measurements. The objective lens reduces the efficiency, likely due to alignment and reflection losses.

We see a typical unmagnified efficiency of $\eta\approx 30\%$. 
The absolute maximum efficiency that can be achieved in a binary diffractive optics system is 50\%, which is limited by the presence of a virtual focus. Higher efficiencies would be obtainable using blazed and/or multi-level diffractive optics \cite{Kedmi1990,ifancharles,binaryblaze}. 
The efficiency values allow us to use the trap widths and depths to estimate the actual trap parameters of our rings. This is again shown in Table~\ref{tab:ringparams} for an illuminating beam power of $10\,$W and an efficiency of $\eta=30\%$.

\begin{figure}[!t]
	\centering
	\includegraphics[width=\textwidth]{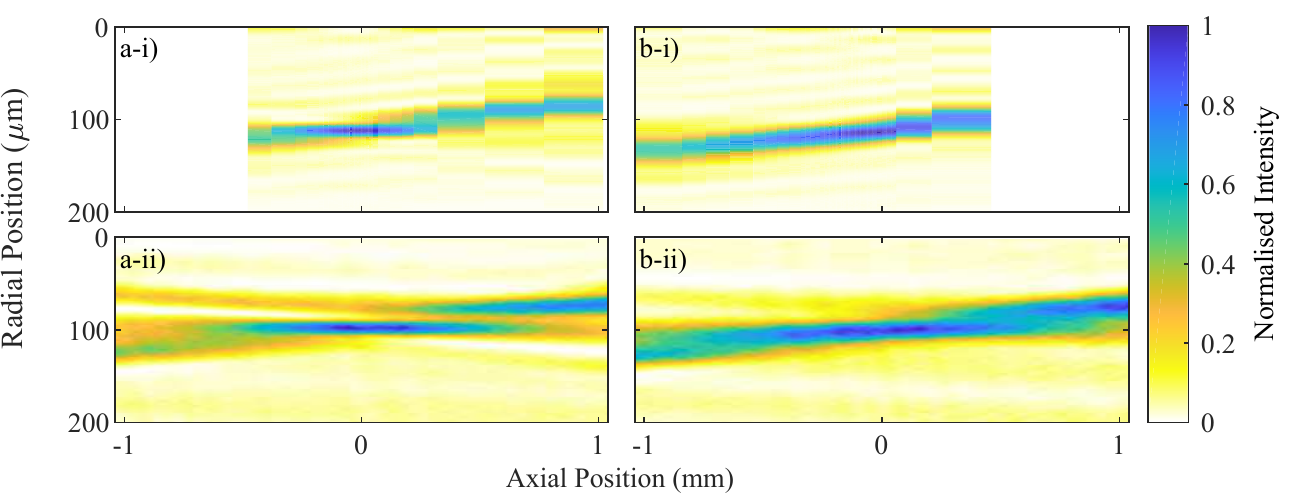}
	\caption[Experimental propagation maps]{\label{fig:prop} Experimental (i) and simulated (ii) propagation of the radial profile of $r_0=100\,\upmu$m, $f=18.2\,$mm rings with (a) $w=5\,\upmu$m  and (b) $w=10\,\upmu$m rings through their focal planes. The axial position is defined such that the FZP is located at $-f$.}
\end{figure}

Axial trapping behaviour is determined by mapping the average radial profile as one scans through the focal plane of the FZP by translating the FZP.  Experimental and simulated intensity maps for $r_0=100\,\upmu$m, $f=18.2\,$mm and  $w=5$ and $10\,\upmu$m are shown in Fig.~\ref{fig:prop}.  
Close to the foci, the rings propagate similarly to a Gaussian beam, whereas further from the focus, additional components can be seen which derive from the binary and diffractive nature of the zone plate. The most obvious
of these diffraction effects are the appearance of `shoulders' on either side of the focus (radially). These features are common to both experimental data and simulations.

By approximating the axial propagation as Gaussian, we obtain experimental $1/e^2$ widths of $8.8(1)\,\upmu$m  and $13.6(1)\,\upmu$m, for the $5\,\upmu$m and $10\,\upmu$m width rings respectively, very close to the values seen transversely in Table~\ref{tab:ringparams}. Note that FZPs with a shorter focal length have the smallest observed ring widths $w$, and due to the scaling laws (Eq.~\ref{scaling}) will provide stronger radial confinement and even stronger axial confinement. The $f=7\,$mm, $r_0=100\,\upmu\textrm{m},$ $w=5\,\upmu\textrm{m}$ ring has a predicted axial trap frequency of $150\,$Hz (Eq.~\ref{nuz}), which is already comparable to the confinement of experiments requiring a separate light sheet for axially trapping \cite{Eckel2014,Marti2015,birkl,boshier2009,hadzibabic2012}.

\subsection{Analysis of trap smoothness} \label{sec:smoothness}


From the fit residuals shown in Fig.~\ref{fig:rings} we can gain a qualitative impression of the smoothness of the traps that the rings provide. A simple quantitative measure of trap `goodness' uses the root mean squared (RMS) error between an ideal trap and the observed trap. This is an established method used in SLM holography as an optimisation parameter for design algorithms \cite{Bruce2015, Bruce2011, Pasienski2008, Gaunt2012} or to compare design methods \cite{Henderson2016}.  We calculate $E$ the RMS error of the top 10\% of the potential, relative to total potential depth, and this is shown in Table~\ref{tab:ringparams}.

\begin{figure}[!t]
	\centering
	\includegraphics[width=\textwidth]{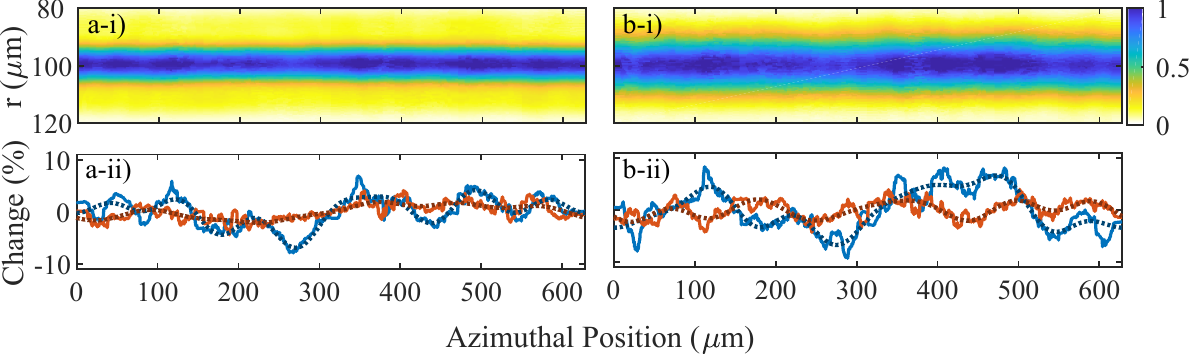}
	\caption[Fourier analysis of roughness]{\label{fig:Roughness} 
Variation in ring parameters for $r_0=100\,\upmu$m, $f=18.2\,$mm rings with $w=5\,\upmu$m (a) and $10\,\upmu$m (b). For each ring we show (i) the unfolded ring, and (ii) the variation in trap depth and width as a percentage of the average value (solid blue and red respectively). The dashed blue and red lines show an $8^\textrm{th}$ order Fourier fit to the variation. Error bars from the slice-fitting routine are too small to be visible and are therefore omitted.}
\end{figure}

In the region of highest intensity (the trap minimum), the magnitude of the angle-averaged RMS residuals ($E$ in Table~\ref{tab:ringparams}) are all below 5\%, with typical peak azimuthal deviations less than 10\% (Fig.~\ref{fig:Roughness}), and the majority of deviations from the target pattern lie outside the trapping region. 
It is difficult to disentangle the origin between errors from the kinoform design and errors that are artefacts of the imaging system. In principle one could begin to study this with full control of the phase and intensity of the incident field. 


The values in Table~\ref{tab:ringparams} give an upper bound on the RMS error, and do not allow for the cartesian-to-polar imaging conversion, shot noise at each camera pixel, pixel-pixel sensitivity variation and imperfect imaging optics/alignment. However, this still suggests all rings have an RMS error below 3-5\%, which is compatible with atom guiding -- with the caveat that large intensity gradients yield acceleration.  The RMS errors in the non-magnified system are higher, probably due to the difficulties in aligning the zone plate in this arrangement, and digitization effects because the ring is then only one or two pixels wide. In future we intend to use ultracold atoms in the guide, as the best probe of the actual local light potential. 

More can also be learnt through Fourier analysis of (`unfolded') azimuthal potentials. By analysing the variation of trap parameters, we can isolate the information most important to the atoms, i.e.\ the magnitude and length scale of roughness. Fig.~\ref{fig:Roughness} (a,b) illustrates two example rings analysed in this way with: (i) the `unfolded' azimuthal (normalised) potential; (ii) the variation in trap depth (blue) and trap width $w$ (red) as a percentage variation from the mean.

This analysis shows that trapping parameters vary between 5-10\%  from their mean values which is similar to the variation limit suggested in previous literature~\cite{Wright2013, Gaunt2012, Bruce2015} and so they are suitable for atom trapping.  The Fourier analysis indicates that low frequency roughness dominates, meaning that the trap parameters tend to vary slowly across the circumference of the ring. This may affect the expansion dynamics causing uneven speed around the two arms of an interferometer, but slow variation should be easily corrected with an SLM \cite{fzp3,Clark2016}.  Many of these defects and roughness may be artefacts of the imaging system as they vary with alignment or are artefacts of the `unwrapping' process, this means that our error measurements constitute a worst-case scenario.

\section{Conclusion}

Fresnel zone plates, in a variety of geometries, were manufactured and optically tested. We used a variety of rings and other atomtronic-type potentials (such as Y-junctions), with the ring potentials currently showing the most promising results -- if the input spatial intensity pattern is Gaussian \cite{fzp3}. 
The rings were then analysed to determine their trapping parameters and the smoothness of the trap. The rings were reproduced without observable ellipticity, with radii corresponding to the design parameters. 
 Imperfect illumination (a $1\,$mm $1/e^2$ radius Gaussian) and numerical aperture limits the widths of the rings produced.

We were able to make benchmarking measurements of the quality (smoothness) of the rings. All rings had an upper limit root mean squared error of 3-5\%, corresponding to  roughness atoms would experience in the potential.  This is expected to be compatible with atom trapping and guiding. Trap parameters ($1/e^2$ radius, trap depth and radii of the rings) varied by a maximum of 10\% amplitude, with the dominant variation occurring in the first few (i.e.\ slowly varying) azimuthal modes.  The benchmarking is limited by aberrations inherent to the imaging system and by the analysis algorithms involved in converting from Cartesian to polar co-ordinates. 

In future we will present the results of a hybrid SLM-FZP system, combining the strengths of both technologies, with a view to ideal next-generation FZPs with enhanced flexibility \cite{fzp3}. 

The dataset associated with this paper is available online \cite{dataset}.

\section*{Funding}
Engineering and Physical Sciences Research Council (EP/M013294/1); Defence Science and Technology Laboratory (DSTLX1000095638R).

\section*{Disclosures}
The authors declare no conflicts of interest.

\section*{Appendix}
\label{appendix}
This appendix provides the numerical relations used to determine experimentally relevant dipole trap parameters \cite{grimm2000,Franke-Arnold2007} and their scaling laws for Table~\ref{tab:ringparams}. We consider the ring spatial intensity (Eq.~\ref{eq:ringeq}) in conjunction with the wavelength of light used and the atomic species under consideration. For example, the potential (as a temperature) is given by $U/k_{B}=-\overline{\alpha} I$ where the effective susceptibility $\overline{\alpha}=\sum_j{\alpha_j ({\Delta_{j+}}^{-1}+{\Delta_{j-}}^{-1})}$ uses $\alpha_j$, the susceptibility of the transition at frequency $\nu_j$, and  $\Delta_{j\pm}=\nu_j\pm\nu_\textrm{D}$ are the sum and difference frequencies between the atomic transition and  dipole laser frequencies, respectively. In rubidium there are two main contributions to the susceptibility from the D$_2$ and D$_1$ lines, and at $1071\,$nm we have $\overline{\alpha}=1.50\times10^{5}\,\upmu\textrm{K}\,(\upmu\textrm{m})^2\,\textrm{W}^{-1}.$ 
By comparing the Taylor expansion of the potential near $r=r_0$ to a harmonic potential, i.e.\ 
\[U-U_0\approx 2 k_B \overline{\alpha} I_0 (r-r_0)^2/w^2= m (2 \pi \nu_r)^2 (r-r_0)^2/2,\] 
where $k_\textrm{B}$ is the Boltzmann constant, $m$ is the atomic mass, $\nu_r$ is the radial trap frequency, and 
\begin{equation}
\label{temp}T=-U_0/k_B= \overline{\alpha} I_0\approx 1.91\times 10^4 \frac{P_\eta}{r_0\, w} \upmu\textrm{K}\,(\upmu\textrm{m})^2\,\textrm{W}^{-1}
\end{equation} is the ring potential minimum as a temperature, we arrive at \begin{equation}\label{nur}\nu_r=\frac{{U_0}^{1/2}}{m^{1/2} \pi w}=\frac{\sqrt{k_B \overline{\alpha} P_\eta}}{(2^{1/2} \pi^{3/2} r_0\,w\, m)^{1/2} \pi w}\approx 430\, \frac{\sqrt{P_\eta}}{{r_0}^{1/2}\,w^{3/2}}\,\textrm{kHz}\,(\upmu\textrm{m})^2\,\textrm{W}^{-1/2}.\end{equation}

To determine the atomic photon scattering rate $R$, and thereby the heating rate in the ring, we note $R\propto I\,{\Delta_j}^{-2}$ and $T\propto I\, {\Delta_j}^{-1}$. At our dipole wavelength for $^{87}$Rb we have the simple law \begin{equation}\label{rate}
R\approx 6.5\times 10^{-3}\,T\,\textrm{Hz}\, (\upmu\textrm{K})^{-1}.\end{equation}  One can also determine the maximum radial acceleration possible in the ring, using $a_\textrm{max}=\max(|dU/dr|),$ which for our Gaussian intensity is maximal at locations $|r-r_0|=\pm w/2,$ leading to a scaling law: \begin{equation}\label{amax}
a_\textrm{max}=\frac{-2 U_0}{e^{1/2} w} = 2.25\times 10^5\,\frac{P_\eta}{r_0\,w^2}\,g \,(\upmu\textrm{m})^3\, \textrm{W}^{-1},\end{equation}
in units of Earth's surface gravitational acceleration  $g=9.81\,\textrm{m}\,\textrm{s}^{-2}.$ If one assumes that the ring focus axially propagates like a Gaussian beam in one dimension (cf.~Fig.~\ref{fig:prop}), i.e.\ $w(z)=w \sqrt{1+(z/z_R)^2}$ where $z_R=\pi\,w^2/\lambda$ is the Rayleigh range, one can show that the axial trap frequency in the ring is then given by: \begin{equation}\label{nuz}
\nu_z=\frac{\lambda}{2\pi w} \nu_r \approx 68.3 \frac{\sqrt{P_\eta}\,\lambda}{{r_0}^{1/2}\,w^{5/2}}\,\textrm{kHz}\,(\upmu\textrm{m})^2\,\textrm{W}^{-1/2}.
\end{equation}
The dramatic scaling with beam waist $w$ indicates that additional axial confinement due to a light sheet could be obviated with a sufficiently radially tight trap. One then needs to ensure the atoms have sufficiently low density to avoid any issues with atomic interactions.


\begin{thebibliography}{99}
\bibitem{Kedmi1990} The etching process is a combination of photolithography and dry plasma etching steps. Patents US5073007A, US5227915A.
\end{thebibliography}
\end{document}